# GEMs with Double Layred Micropattern Electrodes and their Applications

A. Di Mauro, P. Martinengo, E. Nappi, R. Oliveira, V. Peskov, F. Pietropaolo, P. Picchi

*Abstract* We have developed and tested several new designs of GEM detectors with micropattern electrodes manufactured by microelectronic technology. In one design, the inner layer of the detector's electrode consists of thin metallic strips and the outer layer is made of a resistive grid manufactured by a screen printing technology. In other designs, the electrodes were made of metallic strips fed by HV via micro-resistors manufactured by a screen printing technology. Due to these features, the new detectors have several important advantages over conventional GEMs or ordinary thick GEMs. For example, the resistive grid (in the first design) and the screen printed resistors (in other designs) limited the current in case of discharges, making these detectors intrinsically spark-protected.

We will here describe our tests with the photosensitive versions of these detectors (coated with CsI layers) and the efforts of implementing them in several applications. In particular, we will focus on our activity towards the ALICE RICH detector upgrade and on tests of simplified prototypes of cryogenic dark matter detectors.

## I. INTRODUCTION

Hole-type gaseous multiplier [1-4] due to their unique properties offer new possibilities in the detection of photons and charged particles. The most commonly used hole-type detector today is the so-called Gas Electron Multiplier (GEM)[3]. It is accepted in several large-scale experiments at CERN and elsewhere. However, in spite of its great success, the GEM as any other micropattern detector, is a rather fragile device and can be easily damaged by sparks developing at high gains of operation.

The origin of these breakdowns is today well understood. In the case of the poor quality detectors, the discharges are triggered by the presence of microdefects like sharp edges, micro-particles remaining after the production both inside and outside the holes, dirty spots (which are often semiconductive) and so on. In good quality detectors, the breakdowns appear when the total charge in the avalanche reaches some critical value:

$$Q_{crit}=A_{max}n_0 \sim 10^6\text{-}10^7 \text{electrons (1)},$$

where $A_m$ is the maximum achievable gas gain and $n_0$ is the number of primary electrons created by the radiation in the active gas volume of the detector.[*] Thus in the case of the detection of single electrons ($n_0$=1) the $A_{max}$ can be as high as $10^6$. However, in the case of the detection of radiations producing $n_0 \gg 1$ primary electrons, the maximum achievable gain will be reduced. For example, in the case of the detection of x-rays from a $^{55}$Fe radioactive source (each photon creates $n_0 \sim 220$ electrons), the maximum sustainable gain will be $\sim 10^4$ and in the case of alpha particles ($n_0 \sim 10^5$ electrons) the maximum achievable gains will be only $\sim 10^2$.

Hence, if GEMs are used for the detection of single photoelectrons, any radioactive background creating $n_0 > 1$ primary electrons will cause breakdowns. Therefore unfortunately, sparks are unavoidable at operation at high gains. Of course, the GEM community learns how to cope with the sparking problems: they use segmented GEM (to reduce the detector's capacitance), several GEMs operating in cascade (due to the diffusion effect [7] the value of the $Q_{crit}$ increases) and spark-protected electronics. However, the experience in running of the GEM-based PHENIX Hadron blind detector indicates that in spite all efforts the GEMs can still be damaged by sparks [8]. This is why we recently suggested a different approach: spark-proof GEMs with resistive electrodes instead of traditional metallic ones [9, 10]. At low gas gain and low counting rates this detector operate as a usual GEM, however in the case of high gain operations, high counting rates or sparking this detector is more resembling RPCs, for example it is intrinsically spark protected.

We recently introduced a new advanced design of the RETGEM, which combines two approaches: a spark protecting resistive layer and the high segmentation of electrodes allowing one to reduce the capacitance contributing to the discharge power [11]. In this work we present results of comparative studies of several versions of such detectors with the main focus on their possible applications to the detection of single electrons. In particular we are interested in investigating the feasibility of their application to RICHs and

Manuscript received November 14, 2008.

A. Di Mauro is with the PH division, CERN, Geneva-23, Switzerland, CH-1211, (telephone: +4122767-6612, e-mail:antonio.di.mauro@cern.ch).

P. Martinengo is with the PH division, CERN, Geneva-23, Switzerland, CH-1211,(telephone:+4122767-8434,e-mail: paolo.martinengo@cern.ch).

E. Nappi is with INFN Bari, Bari, Italy, (telephone: +390805443202, e-mail: Eugenio.Nappi@ba.infn.it).

R. Oliveira is with TS Div., CERN, Geneva-23, Switzerland, CH-1211, (telephone:+4122767-3745, e-mail: Rui.de.Oliveira@cern.ch).

V. Peskov is with the Institute for Nuclear Research UNAM, Mexico and with the PH Div. of CERN, Geneva-23 (telephone: +4122767-4643, e-mail: vladimir.peskov@cern.ch).

F. Pietropaolo is with INFN Padova, Padova, Italy (telephone: +4122767-5716, e-mail: Francesco.Pietropaolo@cern.ch)

P. Picchi is with INFN Fracati, Frascati, Italy (telephone: +4122767-5716, e-mail: Pio.Picchi@cern.ch)

---

[*] A similar limit was empirically established quite a long time ago by H. Raether [5] for parallel-plate avalanche chambers and is respectively called the "Raether limit". However, it was recently discovered [6] that a similar limit applies for every micropattern detectors: GEMs, MICROMEGAS and others.

## II. DETECTOR DESIGNS AND EXPERMENTAL SETUPS

Three spark-protected GEMs designs were developed and studied in this work. The design of the first detector, called the S-RETGEM, is described in a recent preprint [11]. It has double–layered micropattern electrodes: an inner layer consisting of thin metallic strips and an outer layer comprised of resistive grids manufactured by a screen printing technology on the top of metallic strips. The resistive layers make the detector intrinsically spark-protected. Fig 1 shows a magnified photo of this detector on which one can see the metallic strips with holes inside and a resistive grid on top of them. A peripheral region of this detector has metallic pads to which amplifiers can be connected or a high voltage applied.

The second detector's design was a modified "thick GEM"-TGEM (to know more about TGEMs see [4, 12, 13]). In this design both electrodes consisted of parallel metallic strips each of them containing one row of holes (see Fig. 2) The strip width was 0.7 mm, the hole's diameter was 0.5 mm and their pitch was of 0.8 mm; the detector's thickness was 1 mm and its active area of $3\times 3$ cm$^2$. The strips on one sides of the detector were oriented perpendicular to those located on the opposite side. Each strip was fed with the HV via separate micro-resistors manufactured by a screen printing technology. Such a design (we called it a "strip TGEM" or S-TGEM) even if the strips were not coated with the resistive layers, allows to considerably reduce the sparking energy (due to their low capacitance and resistivity), in this way occasional sparks do not damage the detector.

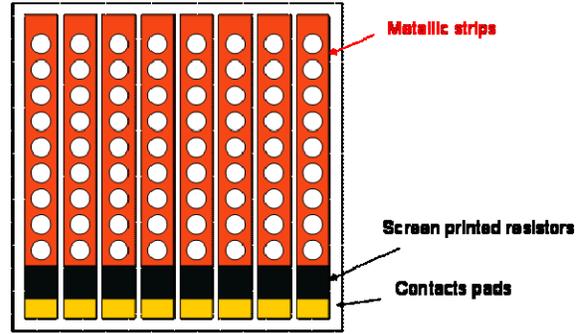

Fig.2. A schematic drawing of the S-TGEM in which each metallic strip contains one row of holes. Strips are fed with HV via micro-resistors manufactured using a screen printing technology.

A similar concept was used in the third design: it has the same geometry as standard GEMs (the hole's diameter was 70 μm, the pitch 140 μm, the detector's thickness 50 μm and the active area 10x10 cm2), but with electrodes made of metallic strips manufactured using a photolithographic technology on the Kapton surface (we called it S-GEM). For simplicity each strip (1 mm in width) contained seven rows of holes . As in the previous design, each strip was connected to the HV electrode via micro-resistors manufactured by a screen printing technology (see Fig. 3). Strips on the opposite sides of the S-GEM were oriented perpendicular to each other.

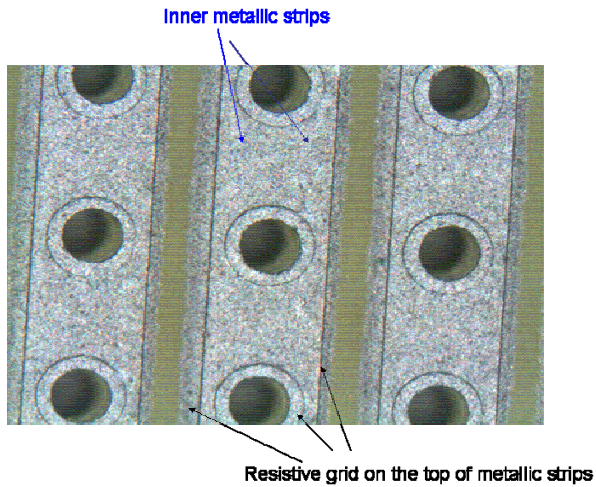

Fig. 1. A magnified photo of the S-RETGEM with holes of 0.3 mm in diameter. The resistive grid and the inner metallic strips are clearly visible.

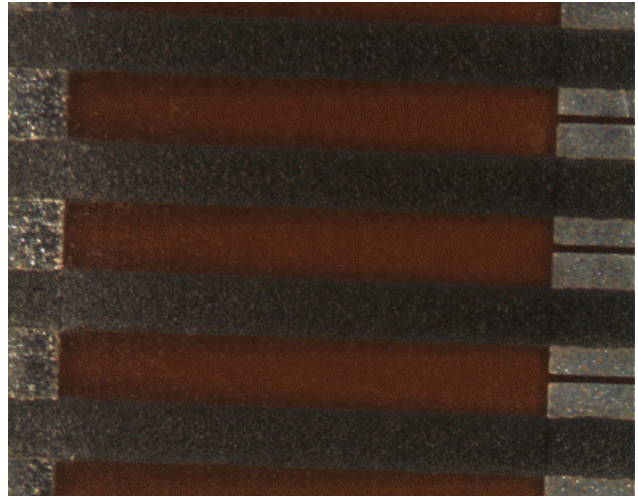

Fig. 3. A magnified photo showing micro-resistors manufactured between the metallic strips (on the right) and a high voltage electrode (on the left).

In several comparative studies small (2x2 cm$^2$ active area) RETGEMs with resistive Kapton electrodes (similar to the ones described in [9]) were used. Usually such RETGEMs have higher qualities than screen printed ones allowing one to achieve almost ten times higher gas gains. One of our new Kapton RETGEM prototypes had also metallic strips manufactured in between the holes on the top of the resistive Kapton electrodes.

Two experimental setups were used in this work. One was dedicated to the study of the strip hole-type detectors for RICH applications whereas the second one was used to study their operation at cryogenic temperatures.

The schematic drawing of the first experimental setup is shown in Fig. 4. It consists of a gas chamber with a $CaF_2$ window, inside which any the mentioned above detectors can be installed, ,a monochromator combined with a Hg lamp and a gas system allowing to pump the chamber or flush it with various gases: Ne, Ar or a mixture of Ne with $CH_4$. In all tests the detectors cathodes were coated with CsI layers (0.35μm thick). In this setup we could measure either the photocurrent from various electrodes, in order to evaluate the gas gain and the photoelectron collection efficiency or the charge signals produced by the avalanches. The procedure followed to measure the quantum efficiency measurements is described in [14, 15].

The second experimental set up is shown in Fig 5. It consists of a specially designed gas chamber allowing the cooling to cryogenic temperature to take place as well as being pumped or flushed with varies gases. Inside the gas chamber a single or double (operating in a cascade mode) hole-type detector can be installed. In the case of low temperature tests the detector was placed inside the dewar filled with liquid nitrogen (78 K) or a mixture of dry ice with alcohol (~195 K) or alcohol with $LN_2$ (~165 K).

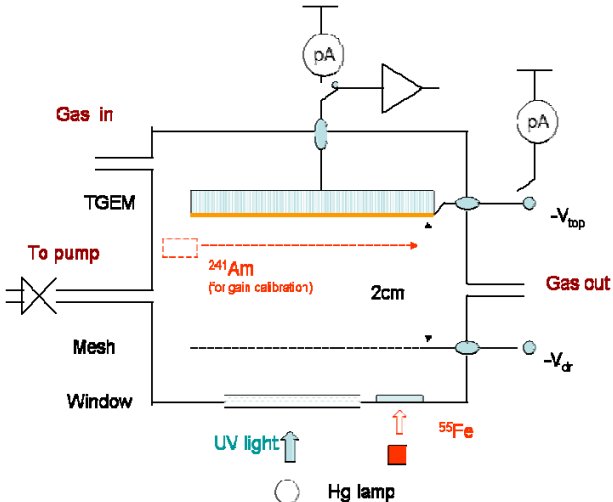

Fig. 4. A schematic drawing of the experimental setup used for the tests at room temperatures.

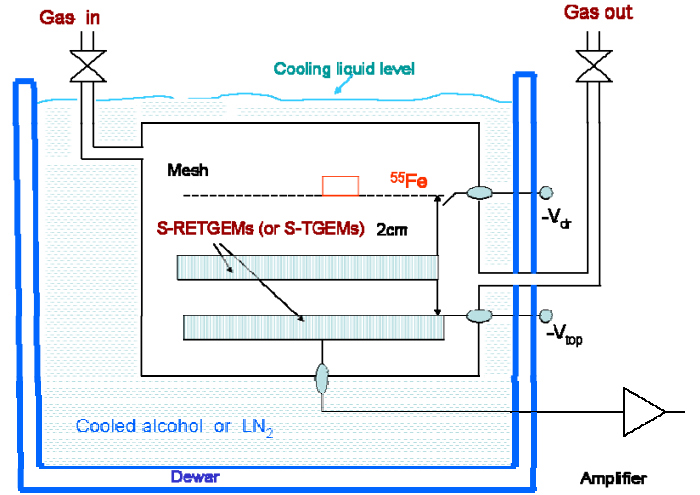

Fig. 5. A schematic drawing of the experimental setup used for the tests at cryogenic temperatures.

### III. RESULTS

The use of strip-type spark-protective GEMs can be attractive in many applications. As an example, we present in this paper some results oriented to their application to RICH and dark-matter noble liquid detectors.

#### A. Test oriented to RICH applications

In the framework of the ALICE experiment at LHC, it has been proposed to build a Very High Momentum Particle Identification Detector (VHMPID) with the aim to upgrade the current ALICE layout [16]. The VHMPID should be able to identify, track-by-track, protons up to 26 GeV/c enabling to study the leading particles composition in jets (correlated with the $\pi^0$ and /or γ energies deposited in the electromagnetic calorimeter).

Due to the very limited space available in the ALICE detector, the VHMPID will be composed by several small (~1x1x1 $m^3$) modules.

The design of the VHMPID module which is presently under study is described in [17]. It will be a focusing- type RICH detector with a gaseous radiator ($CF_4$ or $C_5F_{12}$). The key element of the VHMPID design is a compact planar photodetector. Simulations show that it should have a few mm position resolution and at least 12% quantum efficiency at 185 nm [17]. Because the main task of the VHMPID will be to detect single photoelectrons produced by Cherenkov radiation, it should operate at a gas gain above $10^5$ and thus it will have an elevated risk of sparking. This is why an efficient spark protection of the photodetector is absolutely necessary. Therefore, one of the attractive candidates for the VHMPID photodetector could be a CsI coated hole-type structure protected either with a resistive layer or with in-situ resistors.

In order to choose the most suitable detector for the VHMPID, we performed comparative studies of hole-type detectors described in paragraph II using the setup shown in Fig. 4. In particular, we have measured their maximum achievable gas gains, quantum efficiency and photoelectron collection efficiency.

Fig. 6 shows the gain vs. voltage curves measured with a S-RETGEM for two polarities of the electric field in the drift region: a negative one $E_{dr}$= -250 V/cm and a positive one (inversed polarity) $E_{dr}$=+250 V/cm [11]. The inversion of the electric field in the drift region allows to suppress the contribution of the natural radioactivity and to additionally increase the maximum achievable gains (see the introduction and equation 1). Note that the reversed-field method has already been successfully implemented in cascaded-GEM photon detectors with reflective CsI photocathodes, in order to suppress charged-particle background in high energy physics experiments [18, 19].

From Fig. 6 one can see that at each polarity of the electric field in the drift region, the gas gain achieved in Ne is an order of magnitude higher than in Ar-based gases. Note that in Ar-based mixtures the operational voltages were considerably higher that in Ne. Thus one can speculate that breakdowns in the Ar-based mixture are mainly triggered by the hole imperfections (whereas in Ne-filled detector the Raether limit could be reached).

Fig. 7 shows gain vs. voltage curves measured with a S-TGEM. One can see that in the case of the detection of the UV light, the detector can operate at gains up to $10^6$. We attribute these much higher achievable gain than in the case of S-RETGEM to better quality of production of TGEM compared to screen printed S-RETGEM. It was already mentioned in [10] that the maximum achievable gains of the RETGEMs manufactured by the screen printing technology are lower than the gains achieved with RETGEMs whose electrodes were made of resistive Kapton. One also can see that in Ar the maximum gains achieved in the presence of $^{55}$Fe source were only ~$10^4$, which is consistent with the Raether limit for this gas

Gains achieved with S-GEMs are shown in Fig. 8. As one can see, the maximum achievable gains of our present S-GEM prototypes were quite low, typically below 100. This can also be attributed to the low production quality of the S-GEMs. However, one should take into account that in general the maximum achievable gain of GEMs is usually ~10 times lower than those of TGEMs or Kapton RETGEMs [12]. Thus S-RETGEM and S-TGEM seem more appropriate detectors for the VHMPID.

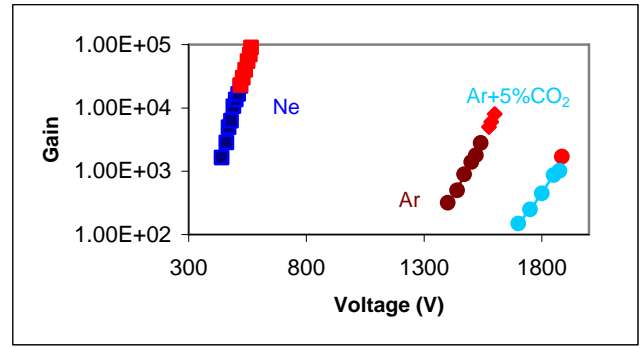

Fig. 6. Gains vs. voltage curves for S-RETGEM measured in Ne, Ar and Ar+5%CH$_4$. Red symbols-results obtained with reversed drift field.

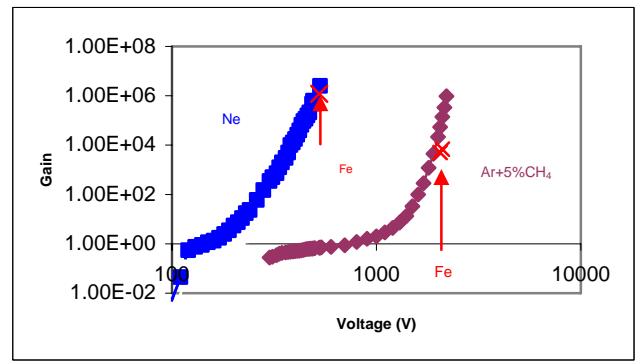

Fig.7. Gains vs. voltage curves for S-TGEM measured with UV light in Ne and Ar+5%CH$_4$. Red crosses indicate the maximum gains achieved with an $^{55}$Fe source. Note that very similar results were earlier obtained with usual TGEM [20].

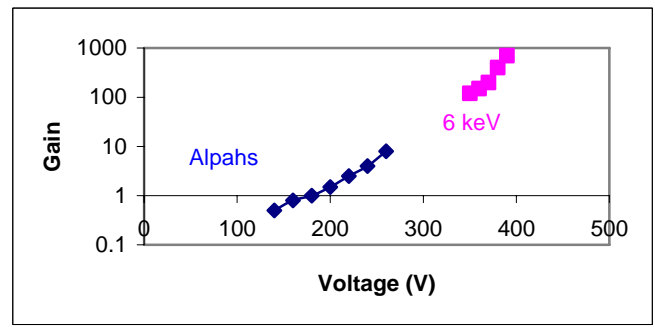

Fig.8. Gains vs. voltage curves for S-GEM measures in Ar.

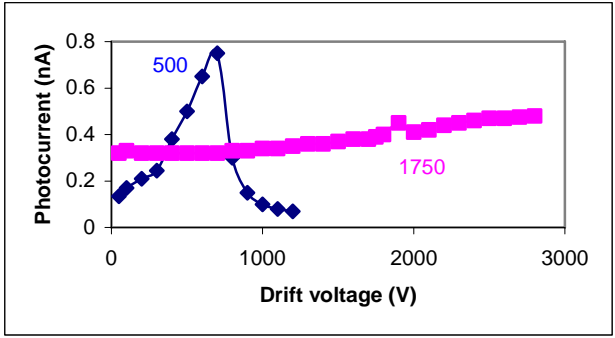

Fig. 9. Photocurrent vs. drift voltage measured with S-TGEM in Ar+5%CH$_4$. The numbers near the curves show the voltages applied across the S-TGEM.

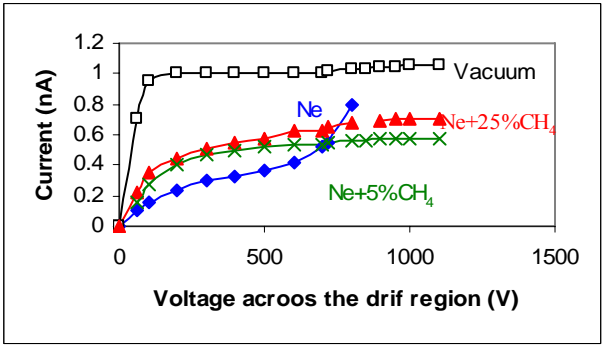

Fig.10. Photocurrent vs. V$_{dr}$ measures in vacuum, Ne, Ne+5%CH$_4$ and Ne 25%CH$_4$. The sharp current rise at V$_{dr}$>650V in Ne is due to the electroluminescence effect. These results confirm measurements reported in [21].

Since it was observed that the polarity of the drift voltage V$_{dr}$ may affect the maximum achievable gain of some detectors especially when operated in Ne (see for example Fig. 6), we performed measurements of the signal amplitudes from the anode of the strip hole-type detectors (at the given voltage across the detector V$_{det}$) as a function of the V$_{dr}$. The obtained results were as follows. Typically, at low values of the V$_{det}$ the anode signal sharply depended on V$_{dr}$ (as an example, see the curve 1 shown in Fig. 9), however with the increase of the V$_{dr}$ the dependence on the drift voltage became weaker and weaker (see Fig. 9). Thus at high gas gains the shape of the gain vs. voltage curves were not sensitive anymore to the drift voltage probably indicating that all photoelectrons created from the CsI photocathode were drifted to the holes (independently of the V$_{dr}$), multiplied there and collected an the anode electrodes.

To confirm this, in the next set of experiments, we measured the photoelectron collection efficiency and quantum efficiency of S-RETGEM and S-TGEM.

Results obtained with both of these detectors were identical so for simplicity we will present below only the experimental data for the S-TGEM. Fig. 10 shows photocurrents (produced by a Hg lamp) and measured on the drift's electrode (see Fig. 4) as a function of the V$_{dr}$ under conditions when the chamber was pumped or filled with various gases. In the case of vacuum, the photocurrent reached saturated value I$_{vac}$ at V$_{dr}$≥100 V. In the gas however, due to the well known back diffusion effect [22], the photocurrent rather slowly increased with the V$_{dr}$. For example, in the case of Ne, the photocurrent reached ~50% of the vacuum level at V$_{dr}$=800 V indicating that only 50% of photoelectrons were extracted from the CsI photocathode at this voltage. In mixtures of Ne with CH$_4$ the extraction efficiency was higher: ~60% for Ne+5%CH$_4$ and ~70% for Ne+25%CH$_4$ at V$_{dr}$ =1kV.

The quantum efficiency of the T-GEM can be defined as:

$$Q(\lambda)=Q_{vac}(\lambda)k(V_{dr}, V_{det}) \varepsilon(V_{dr}, V_{det}) \quad (2),$$

where Q$_{vac}$ is the quantum efficiency of its CsI photocathode measured in the vacuum at a wavelength λ, k is the extraction efficiency in the gas and ε is the collection efficiency of the extracted photoelectrons in detector's holes.

The Q$_{vac}$ was evaluated with respect to the quantum efficiency of the TMAE vapors Q$_{TMAE}$(λ) as it was already done in [15]. For the same geometrical arrangement of the light source and the detectors:

$$Q_{vac}(\lambda)=I_{vac}(\lambda)I_{TMAE}(\lambda) Q_{TMAE}(\lambda) \quad (3),$$

where I$_{TMAE}$ (λ) is a photocurrent value in the reference TMAE detector. Calculations from formula (3) show that the measured current I$_{vac}$=1 nA (see Fig. 10) corresponds to the Q$_{vac}$=17.8% at 185 nm and respectively photocurrent values: 0.5 nA, 0.6 nA and 0.7 nA- measured in Ne, Ne+5%CH$_4$ and Ne+15%CH$_4$ correspond to quantum efficiency values of 8.9% ,10.7% and , 12.4%.

Because Q depends on V$_{dr}$ and V$_{det}$, we also performed quantum efficiency measurements at high gas gains (in this case in counting mode -see [14] for details). We obtained at λ=185 nm in Ne, Ne+5%CH$_4$ and Ne+15%CH$_4$ the following values for the quantum efficiency: 12.3%, 13.8% and 14.7% respectively.

As one can see, the quantum efficiencies measured in the counting mode were typically 20-30% higher than values obtained form the current measurements. This is probably due to the fact that at high V$_{det}$ the electric filed on the top detector's electrode may reach values of ~10 Kv/cm [13] and at such high electric field the coefficient k may approach a value close to unity.

The main conclusion from these studies is that S-TGEMs and S-RETGEMs are promising candidates for the VHMPID: they can operate at high gas gains and have sufficiently high quantum efficiency.

*B. Test oriented for dark matter detectors*

At present, several groups are considering the use of hole-type gaseous multipliers for the detection of the UV light and primary electrons produced by recoils in noble liquid dark matter detectors. For example, we have earlier demonstrated that a CsI coated TGEM is a very robust detector capable of operating stably at cryogenic temperatures up to 78 K [23] (as recently confirmed by the Novosibirsk group [24]). This group also investigated the operation of the old version of the resistive GEM (described in [10]) and fully confirmed our results obtained with this detector at room temperature. However, a strong charging up effect was observed at 78 K.

Because the detectors described in this paper have either different designs (strips instead of conventional unsegmented electrodes) or in some cases were made of a different material (Kapton), their behavior at low temperatures is expected to be more stable. Our preliminary measurements performed with the setup shown in Fig. 5, fully support this assumption. Fig. 11 and 12 show the gain curves measured for S-TGEM at 300K and 88K and for S-RETGEM in Ne and Ar at 300K, 165K and 78K. In Fig. 13 results are presented on the stability measurements at these temperatures. It is evident that the S-RETGEM does not exhibit any strong charging up effect. Similar stability was observed with the S-TGEM.

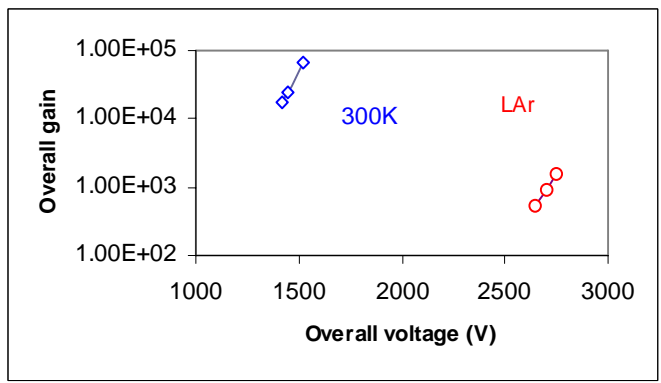

Fig. 11. Gain vs. voltage curves measured with double S-TGEM in Ar at 300K and 78K.

A brief test was done with the Kapton RETGEM; some results of gain measurements are presented in Fig. 14. In contrast to the screen printed RETGEM (tested in [24]) it also exhibited a more stable behavior - see Fig. 15.

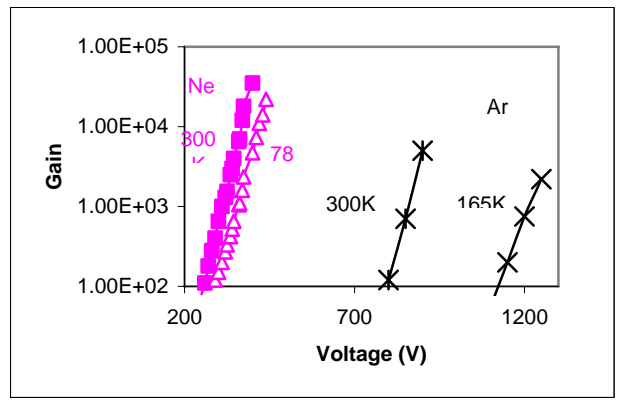

Fig.12. Gains vs. voltage curves for single S-RETGEM measures in Ne at 300 and 78K and in Ar at 300K and 165K.

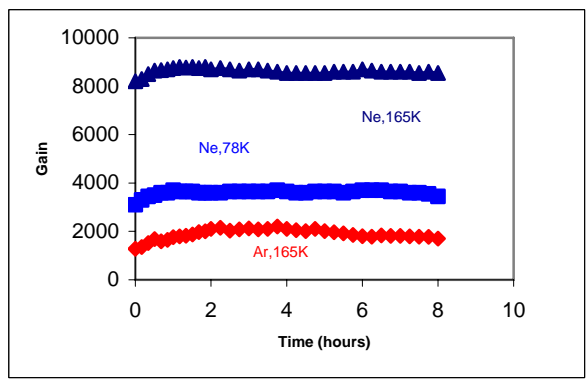

Fig.13. Gain vs. time measured with S-RETGEM at various temperatures in Ne and Ar. Counting rate was about 20Hz/cm$^2$.

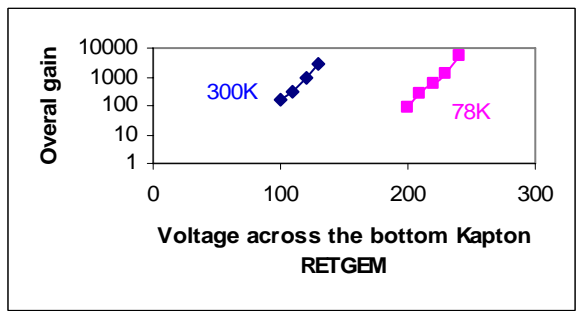

Fig.14. Gain vs. voltage $V_b$ across the bottom of the Kapton S-RETGEM. The voltage on the resistor divider which fed the drift electrode and a top S-RETGEM was kept $V_{total}=3V_b$.

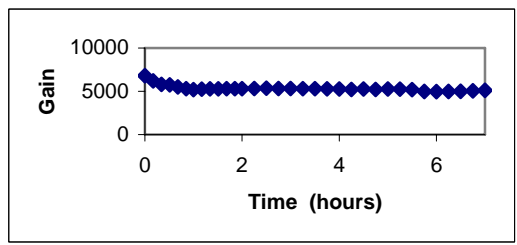

Fig. 15. Gain vs. time measured with double Kapton RETGEM in Ne at 78K

Thus preliminary measurements indicate that all these detector designs could be attractive for noble liquid dark matter detectors.

IV. CONCLUSIONS

Hole-type detectors with strip electrodes have several advantages over conventional designs:

1) Due to the low strip capacitance, the energy released in the sparks is lower than in traditional designs of GEM-like detectors (with unsegmented electrodes). Resistive layers (or in-situ resistors) allow additional strong suppression of the spark's energy to take place.

2) The strip design offers the possibility of taking position information directly from the strips, which can be a convenient option in some applications.

3) In the case of localized defects which triggers breakdowns, the corresponding strip can be disconnected and the detector still operated.

4) Because the strip hole-type detectors can reliably operate at high gas gains with sufficiently high quantum efficiencies, they are attractive candidates for the ALICE VHMPID.

4) Preliminary measurements indicate that strip hole-type detectors allow rather stably operate at cryogenic temperatures and thus also can be useful in such applications as dark matter detectors or noble liquid TPCs.

This is why we believe that strip hole–type detectors may have a great future.

ACKNOWLEDGMENTS

We thank J. Van Beelen and M. Van Stenis for the technical support throughout this work

REFERENCES

[1] A. Del Guerra, M. Conti, G. Gorini et al., "High resistance lead glass tubing for RICH counters and electromagnetic calorimeters," Nucl. Instrum. and Methods, vol. A257, pp. 609-613, 1987
[2] H. Sakurai, T. Tamura, S. Gunji et al.,"A new proportional counter using a capillary plate" Nucl. Instrum. and Methods, vol. A374, pp. 341-344, 1996
[3] F. Sauli, "GEM: a new concept for electron multiplication in gas detectors"., Nucl. Instrum. and Methods, vol. A386, pp. 531-534, 1997
[4] L. Periale, V. Peskov, P. Carlson et al, "Detection of the primary scintillation light from dense Ar, Kr and Xe with novel photosensitive gaseous detectors, " Nucl. Instrum. and Meth., vol. A478, pp. 377-383, 2002.
[5] H. Raether, *Electron avalanches and breakdown in gases.* London, Buterworth, 1964.
[6] V. Peskov, P. Fonte , M. Danielsson et al., "The study and optimization of new micropattern gaseous detectors for high -rate applications," IEEE Nucl. Sci., vol. 48, pp.1070-1074, 2001
[7] P. Fonte, V. Peskov, B. Ramsey, "A study of breakdown limit in microstrip gas counters with preamplification stuctures", Nucl. Instrum. and Methods, vol. A416, pp. 23-31, 1998
[8] W. Amderson et al., "Construction, commissioning and performance of a hadron blind detector for the Phenix experiment at RHIC" Report at the Nuclear Science Sympos., Hawaii, USA, 2007
[9] R. Oliviera, V. Peskov, P. Peitropaolo et al., "First tests of thick GEM with electrodes made of resistive kapton," Nucl. Instrum. and Methods, vol. A576, pp. 362-366, 2007
[10] G. Agocs, B. Clark, P. Martinengo et al., "Developments and the preliminary tests of resistive GEMs manufactured by a screen printing technology," JINST, 3 P02012, 2008, pp.1-10
[11] V. Peskov, P. Martinengo, E. Nappi et al., "Progress in the development of photsensitive GEMs with resistivelectrodes manufactured by a screen printing technology", Preprint arXiv:0807.2718, 2008, pp.1-9
[12] J. Ostling, A. Brahme, M. Danilesson et al., "Study of hole-type gas multiplication structures for portal imaging and other high count rate applications", IEEE Trans. Nucl. Sci, vol. 50 (4), pp.809-819, Aug. 2003.
[13] C. Shalem, R. Chechik, A. Breskin et al., "Advances in Thick GEM-like gaseous electron multiplier-Part I: atmospheric pressure operation", Nucl. Instrum . Meth., vol. A558, pp. 475-489, 2006.
[14] A.G. Agocs, A. Di Mauro, A. Ben David et al., "Study of GEM-like detectors with resistive electrodes for RICH applications", Nucl. Instrum. and Methods, vol. A595, pp. 128-130, 2008
[15] G. Charpak, P. Benaben, P. Breil et al., "Development of a new hole-type avalanche detector and the forst results of their applications" IEEE Trans. Nucl. Sci, vol. 55 (3), pp.1657-1663, June. 2008.
[16] G. Paic, Report at ALICE Physics forum "Very high momentum identification at ALICE- a possibility" http://indico.cern.ch/conferenceDisplay.py?confId=26371,2008
[17] G. Volpe, D. Di Bari, A. Di Mauro et al., "Gas Cherenkov detectors for high momentum charged particle identification in the ALICE experiment at LHC", Nucl. Instrum. and Methods, vol. A595, pp. 40-43, 2008
[18] D. Mormann, A. Breskin, R. Chechik etal., "Operation principles and properties of the multi-GEM gaseous photomultiplier with reflective photocathode," Nucl. Instrum. and Methods, vol. A530, pp. 258-274, 2004
[19] Z. Fraenkel, A. Kozlov, M. Naglis et al., "A hadron blind detector for PHENIX experiment ar RHIC," Nucl. Instrum. and Methods, vol. A546, pp. 466-480, 2005
[20] A. Breskin , V. Peskov, J. Miyamoto et al "New results on THGEM operation" report at the 2d RD51 Collaboration meeteing, Paris, October 2008, http://indico.cern.ch/conferenceDisplay.py?confId=35172
[21] A. Breskin, V. Peskov, "TGEM gain investigation:UV vs. X-rays and photoelectron extraction", presented at the Gaseous Detectors group seminar at CERN, August 2008
[22] V. Peskov, "Secondary processes in gas-filled counters-part II," Sov. Phys. Tech Phys, 22(3), pp.335-338, 1977
[23] L. Periale, V. Peskov, C. Iacobaeus et al., "A study of the operation of especially designed photosensitive gaseous detectors at cryogenic temperatures," Nucl. Instrum. and Methods, vol. A567, pp. 381-385, 2006
[24] A. Bondar, A. Buzulutskov, A. Gerbenuk et al., "Thisk GEM versus thin GEM in two-phase argon avalanche detectors, "JINST 3 P07001, pp1-20, 2008